\documentclass[10pt]{bmc_article}

\usepackage{cite} % Make references as [1-4], not [1,2,3,4]
\usepackage{url}  % Formatting web addresses  
\usepackage{ifthen}  % Conditional 
\usepackage{multicol}   %Columns
\usepackage[utf8]{inputenc} %unicode support
\usepackage{graphicx}% Include Fig. files
\usepackage{dcolumn}% Align table columns on decimal point
\usepackage{bm}% bold math
\usepackage{textcomp}
\usepackage{revsymb}
\usepackage{amssymb}
\usepackage{ifsym}
\usepackage{wasysym}

\newboolean{publ}

\newenvironment{bmcformat}{\baselineskip20pt\sloppy\setboolean{publ}{false}}{\baselineskip20pt\sloppy}

\begin{document}
\begin{bmcformat}


\title{Metal silicide/poly-Si Schottky diodes for uncooled\\ microbolometers}
%%%%%%%%%%%%%%%%%%%%%%%%%%%%%%%%%%%%%%%%%%%%%%
%%                                          %%
%% Enter the authors here                   %%
%%                                          %%
%% Ensure \and is entered between all but   %%
%% the last two authors. This will be       %%
%% replaced by a comma in the final article %%
%%                                          %%
%% Ensure there are no trailing spaces at   %% 
%% the ends of the lines                    %%     	
%%                                          %%
%%%%%%%%%%%%%%%%%%%%%%%%%%%%%%%%%%%%%%%%%%%%%%

\author{Kirill~V~Chizh\correspondingauthor$^1$\and
\email{Kirill~V~Chizh\correspondingauthor - chizh@kapella.gpi.ru}
Valery~A~Chapnin$^1$\and
\email{Valery~A~Chapnin - chapnin@kapella.gpi.ru}
Victor~P~Kalinushkin$^1$$^,$$^2$\and
\email{Victor~P~Kalinushkin - vkalin@kapella.gpi.ru}
Vladimir~Y~Resnik$^1$\and
\email{Vladimir~Y~Resnik - vjresnick@gmail.com}
Mikhail~S~Storozhevykh$^1$
\email{Mikhail~S~Storozhevykh - storozhevykh@kapella.gpi.ru}
and             Vladimir~A~Yuryev\correspondingauthor$^1$$^,$$^2$%
         \email{Vladimir~A~Yuryev\correspondingauthor - vyuryev@kapella.gpi.ru}
%\homepage{http://www.gpi.ru/eng/staff\_s.php?eng=1\&id=125}
      %
      }

%%%%%%%%%%%%%%%%%%%%%%%%%%%%%%%%%%%%%%%%%%%%%%
%%                                          %%
%% Enter the authors' addresses here        %%
%%                                          %%
%%%%%%%%%%%%%%%%%%%%%%%%%%%%%%%%%%%%%%%%%%%%%%

\address{%
    \iid(1)A\,M\,Prokhorov General Physics Institute of the Russian Academy of Sciences, 38 Vavilov Street, Moscow, 119991, Russia\\
    \iid(2)Also at Technopark of GPI RAS, Moscow, 119991, Russia
}%

%%% PACS numbers
%\pacs{73.30.+y, 73.40.Ns, 85.60.Bt, 85.60.Gz, 07.57.Kp, 73.50.Lw}

\maketitle

%%%%%%%%%%%%%%%%%%%%%%%%%%%%%%%%%%%%%%%%%%%%%%
%%                                          %%
%% The Abstract begins here                 %%
%%                                          %%  
%% Please refer to the Instructions for     %%
%% authors on http://www.biomedcentral.com  %%
%% and include the section headings         %%
%% accordingly for your article type.       %%   
%%                                          %%
%%%%%%%%%%%%%%%%%%%%%%%%%%%%%%%%%%%%%%%%%%%%%%

\begin{abstract}
        % Do not use inserted blank lines (ie \\) until main body of text.
Nickel silicide Schottky diodes formed on polycrystalline Si$\langle$P$\rangle$ films are proposed as temperature sensors of monolithic uncooled microbolometer IR focal plane arrays. Structure and composition of nickel silicide/polycrystalline silicon films synthesized in a low-temperature process are examined by means of transmission electron microscopy. 
The Ni silicide is identified as multi-phase compound composed by 20 to 40\% of Ni$_3$Si, 30 to 60\% of Ni$_2$Si and 10 to 30\% of NiSi with probable minor content of NiSi$_2$ at the silicide/poly-Si interface. 
Rectification ratios of the Schottky diodes vary from $\sim$\,100 to  $\sim$\,20 for the temperature increasing from 22 to 70{\textcelsius}; they exceed 1000 at 80\,K.
 A barrier of $\sim 0.95$~eV is found to control the photovoltage spectra at room temperature. A set of barriers  is observed in photo-emf spectra at 80\,K and attributed to the Ni-silicide/poly-Si interface. Absolute values of temperature coefficients of voltage and current are found to vary from 0.3 to 0.6\,\%/{\textcelsius} for forward biasing and around 2.5\,\%/{\textcelsius} for reverse biasing of the diodes.       
\end{abstract}

\ifthenelse{\boolean{publ}}{\begin{multicols}{2}}{}

%%%%%%%%%%%%%%%%%%%%%%%%%%%%%%%%%%%%%%%%%%%%%%
%%                                          %%
%% The Main Body begins here                %%
%%                                          %%
%% Please refer to the instructions for     %%
%% authors on:                              %%
%% http://www.biomedcentral.com/info/authors%%
%% and include the section headings         %%
%% accordingly for your article type.       %% 
%%                                          %%
%% See the Results and Discussion section   %%
%% for details on how to create sub-sections%%
%%                                          %%
%% use \cite{...} to cite references        %%
%%  \cite{koon} and                         %%
%%  \cite{oreg,khar,zvai,xjon,schn,pond}    %%
%%  \nocite{smith,marg,hunn,advi,koha,mouse}%%
%%                                          %%
%%%%%%%%%%%%%%%%%%%%%%%%%%%%%%%%%%%%%%%%%%%%%%

%%%%%%%%%%%%%%%%
%% Background %%
%%
%\section*{Content}

\section*{Introduction}

Lately, outstanding achievements have been made in development of a novel class of uncooled microbolometer IR focal plane arrays (FPAs), the ones based on Si-on-insulator diodes as temperature sensors, whose format has reached 2 megapixels with a noise equivalent temperature difference (NETD) of  60 mK at  the frame rate of 15 Hz and the f-number of 1; the same group has also demonstrated a VGA FPA with outstanding NETD of 21 mK (at f/1, 30 Hz) [see, e.\,g., Ref.\,\cite{Diode_bolometers-Mitsubishi-camera-2Mpix} and earlier articles cited therein]. This success, as well as previous achievements in this field \cite{Diode_bolometers-Mitsubishi-technology, Diode_bolometers-Mitsubishi-22mK-camera, Diode_bolometers_SOI-review}, stimulates search for simple CMOS compatible technological solutions based on diode bolometers which would be suitable for mass production of IR FPAs with low cost and NETD figures sufficient for many civil applications \cite{Diode_bolometers-turkish1, Diode_bolometers-turkish2, Diode_bolometers-1, patent-Schottky-bolometer, patent-diode-bolometer}. One of such solutions consists in utilization of  metal/poly-Si Schottky barriers for formation of sets of temperature sensors on bolometer membranes \cite{patent-Schottky-bolometer, Orion-2012}. Schottky barrier bolometer arrays seem to be first  proposed theoretically for very sensitive cooled bolometers \cite{Schottky-Diode_bolometers-cooled}. In this article, nickel silicide Schottky diodes formed on polycrystalline Si$\langle$P$\rangle$ films are proposed as thermosensitive elements of monolithic uncooled microbolometer IR FPAs. 
The possibility of integration of technological process of the silicide based Schottky diode structure formation into the standard CMOS technology of VLSI manufacturing \cite{PtSi-IrSi_Micelec-our-RU} as well as the possibility of cascade connection of Schottky diodes to increase the temperature sensitivity of bolometer elements of FPA and the use of layers of the diode structures as absorbing coatings in bolometers are advantages of these structures.

%%%%%%%%%%%%%%%%%%
\section*{Sample preparation and characterization techniques}  

Schottky barriers were formed on commercial single-crystalline CZ silicon wafers ($\rho = 12~\Omega$cm, (100), $p$-type) coated by a $\sim$\,600 nm thick layer of SiO$_2$ formed by thermal oxidation and a  $\sim$\,180\,nm thick layer of pyrolytic Si$_3$N$_4$ (the dielectric layers simulated a design of the supporting membranes of the previously tested bolometer cells \cite{SiGe-TSR_Micelec, Orion-2012, SiGe_bolometers-voitsekh}). Films of polycrystalline Si$\langle$P$\rangle$ with the thicknesses of $\sim$\,150\,nm were deposited by thermal decomposition of monosilane at the substrate temperature $T_{\rm s}\approx 620${\textcelsius}; then they were  doped by phosphorus by ion implantation ($E$\,=\,35\,keV) to the dose of $5\times 10^{15}$\,cm$^{-2}$  and annealed at 700{\textcelsius} for 30~min. After wafer cleaning in a boiling ammonia-peroxide mixture solution (NH$_4$OH:H$_2$O$_2$:H$_2$O\,=\,1:1:4, 10~min) and surface hydrogenation (HF:H$_2$O\,=\,1:10, 30~sec at room temperature), 
Ni-silicide/poly-Si Schottky diodes were formed by thermal deposition of a  nickel film ($\sim$\,45\,nm thick, $T_{\rm s}\sim 300$\,K, the residual gas pressure $P_{\rm r} <10^{-6}$\,Torr) from a tungsten crucible followed by annealing at 400{\textcelsius} in nitrogen for 30~min. Al contacts to poly-Si were formed by thermal deposition from tungsten crucible in vacuum ($P_{\rm r} < 10^{-6}$\,Torr, $T_{\rm s}\sim 300$\,K) and annealing at 450{\textcelsius} in nitrogen for 15~min. Aluminium contacts to the structure top were deposited in the same way but without annealing. Golden wires were welded to the contact pads. 
Structural perfectness and chemical composition of the layers were explored by means of transmission electron microscopy (TEM).
Test elements for electrical measurements were formed by contact lithography and had the sizes of $\sim 1$\,mm. {\it I--V} characteristics of the Schottky diodes were measured in darkness at different temperatures varied in the range from 20 to 70{\textcelsius} and at the temperature of 80\,K. Photovoltage ($U_{\rm emf}$) spectra  were obtained as described in Ref.\,\cite{VCIAN-2012}; for each photon energy ($h\nu$) the photoresponse value $U_{\rm emf}$ was normalised to the number of incident photons.
Uncoated satellites were used for measurement of sheet resistance ($\rho_{\rm s}$)  of the poly-Si films.
The WSxM software \cite{WSxM} was used for TEM image processing.

\section*{Results and discussion}

A typical TEM micrograph  of the resultant structure (Fig.\,\ref{fig:TEM}) represents images of polycrystalline Ni silicide and polysilicon layers between Si$_3$N$_4$ and Al films.  The   Ni silicide film is seen to be composed by a number  of phases: at least two phases   with the grains close in sizes and comparable volume fractions are distinctly observed by TEM. Bright inclusions  are also observed at Ni-silicide/poly-Si interface; we presumably interpret them as residual silicon oxide particles.

It is also seen  in Fig.\,\ref{fig:TEM} that after formation of the Ni-silicide/poly-Si film, the average thicknesses of the Ni silicide and poly-Si layers became 60 and 135~nm, respectively. Using the mass conservation law this allows us to estimate the density of the silicide film as $\sim$\,7 g/cm$^3$ (we adopt the density of poly-Si to be 2.33 g/cm$^3$ and the density of the initial poly-Ni film to be 8.9 g/cm$^3$). This in turn allows us  to roughly evaluate  the composition of the silicide layer  (the required densities of Ni silicides can be found, e.\,g., in Refs.\,\cite{Murarka, Silicides}). If we postulate that the silicide film consists of only two phases, as it is stated in Ref.\,\cite{Murarka},  then they might be Ni$_2$Si and NiSi (the process temperature did not exceed 450{\textcelsius} and mainly was 400{\textcelsius} or lower; it is known however that NiSi$_2$---or slightly more silicon rich compound \cite{Ni-Silicide_Schottky_barrier_height}, according to Ref.\,\cite{Formation&Stability-Silicides/Poly-Si}, Ni$_{1.04}$Si$_{1.93}$---forms at $T_{\rm s} > 600${\textcelsius} (or even $> 700${\textcelsius} \cite{Silicides}), whereas NiSi and Ni$_2$Si form at $T_{\rm s} > 400$ and 200{\textcelsius}, respectively \cite{Sze, Ni-Silicide_Schottky_barrier_height, Formation&Stability-Silicides/Poly-Si}; according to Ref.\,\cite{Murarka}, appearance of these two low-temperature phases of Ni silicides after annealing in vacuum would evidence that the original Ni film has been completely (or nearly completely) consumed by the growing Ni$_2$Si phase).$^{\rm a}$
In this case the volume fraction of Ni$_2$Si:NiSi\,$\agt$\,85:15 (taking into account all uncertainties, the maximum estimate yields 100\% of Ni$_2$Si); the mass fraction of  Ni$_2$Si  exceeds 88\%. 
This obviously contradicts our TEM observations and makes us assume the presence of the heaviest of the Ni silicides, Ni$_3$Si \cite{Silicides}, which also may form at low temperatures, especially taking into account the possible presence of oxygen in the metal film that, according to Refs.\,\cite{Murarka, Metal-silicide_oxygen}, impedes diffusion of Ni atoms to Ni/Ni$_2$Si interface and, in our opinion, may result in simultaneous formation of Ni$_2$Si and Ni$_3$Si phases in the silicide film. If our assumption is true the silicide film might be composed,  by a rough estimate, by 20 to 40\% of Ni$_3$Si, 30 to 60\% of Ni$_2$Si and 10 to 30\% of NiSi in respective proportions to give  a total of 100\% of the silicide film volume. The lightest (the least dense) silicide phase having a Si-rich stoechiometry (disilicide) may also be available in the form of a thin diffusion layer at the Ni-silicide/poly-Si interface  (this does not contradict our observations) \cite{Ni-Silicide_diffusion_layer}; it may affect the barrier height of  the whole silicide layer, however \cite{Ni-Silicide_Schottky_barrier_height}.

$I$--$V$ characteristics of the structures (Fig.\,\ref{fig:I-V-300K}\,a,\,b) with low-resistance poly-Si ($\rho_{\rm s} \approx 270\,\Omega/\Square$), which forms in our process, manifest a diode behavior with  the rectification ratios changing from $\sim$\,100 to  $\sim$\,20 for the temperature varied from 22 to 70{\textcelsius} (Fig.\,\ref{fig:I-V-300K}\,c). 
At liquid nitrogen temperature the rectification becomes more pronounced and exceeds 1000 at biases exceeding 2~V (Fig.\,\ref{fig:I-V-80K}). It should be noticed that at forward biasing the negative lead was set on the silicide top contact pad whereas the positive one was set on the contact pad to the polysilicon film.  

Photo-emf spectra obtained at 300 and 80\,K (Fig.\,\ref{fig:emf}) demonstrate photoresponse for photons with energies greater and less than the Si bandgap width ($E_{\rm g}$) as well as the presence of a number of potential barriers in the diode film.
Room temperature measurements with and without a silicon filter have revealed the only barrier with the height $\Phi_{\rm rt}\approx 0.95$\,eV (remark that the negative pole of the photodiode was on the contact pad to the silicide  when the diode was  illuminated by the white light). A richer collection of barriers has been revealed at 80\,K. 
The highest one nearly coincides in   energy with $E_{\rm g}$ ($\Phi_0\approx 1.1$\,eV  with 95\,\% confidence limits of 1.08 and 1.14\,eV). 
A lower one $\Phi_1 \approx 0.74$ (with the 95\% confidence limits of 0.66 and 0.78\,eV) is close to the values ascribed in the literature to all Ni silicide barriers with $n$-type Si \cite{Murarka, Sze, Ni-Silicide_Schottky_barrier_height} (equality of barrier heights of all nickel silicides was explained by the presence of similar diffusion layers in all nickel-silicide/silicon interfaces \cite{Ni-Silicide_Schottky_barrier_height}). 
Estimation of the lowest one yields a figure of $\Phi_2\approx 0.51$\,eV (the 95\% confidence band is from 0.48 to 0.54\,eV); a barrier of this height, to our knowledge, has never been connected  with a Ni-silicide/Si transition in the literature.$^{\rm b}$ However, we attribute all the above barriers to the Ni-silicide/poly-Si interface. Our reasoning is as follows. The band structure of a polysilicon film is known to be spatially inhomogeneous: A strong potential relief is associated with grain boundaries \cite{Matare}. In  $n$-Si, even in the heavily doped $n^+$ one, there may exist depleted or even $p$-type spatial domains \cite{Matare} which, on the one hand, as a result of band-to-band transitions may be sources of electron-hole pairs  which, in turn,  are separated by the potential relief and generate the photo-emf of the observed polarity because, despite that the potential peaks should be more or less symmetrical and the electron-hole pairs should arise with close likelihoods on both their slopes, a part of electrons escapes from the Si film accumulating in silicide whereas holes are localized at the grain boundaries. This process may give rise to the photovoltage under irradiation by photons with energies   $h\nu \apprge E_{\rm g}$. In addition to charge separation on opposite sides of the film, this process also increases the potential relief. On the other hand, grain boundaries may serve as potential barriers for electrons localized in $n^+$-Si grains segregating them from the Ni-silicide film and  producing the photo-emf   of the observed polarity due to electron injection into the silicide under the effect of photons with  $h\nu < E_{\rm g}$.
$n^+$-Si potential valleys adjoining the Ni-silicide film form ohmic contacts. This argumentation explains the presence of the only barrier $\Phi_{\rm rt}$ detected at room temperature as well as the observed polarity of both the resultant photovoltage and the forward current. 

A model of processes taking place at liquid nitrogen temperature is some more tricky. As it follows from the $I$--$V$ characteristics (Fig.\,\ref{fig:I-V-80K}), the free electrons are partially frozen out in the structure and the Fermi level  moves down that increases the barriers for electron photoinjection into the silicide film. It makes $\Phi_{\rm rt}$ move to the  right in energy to appear in the photovoltage spectra as $\Phi_0$. Two processes can be mixed in this conditions, band-to-band transition with separation of electron-hole pairs and electron injection into the silicide over the potential barrier, both generating photo-emf.  
In addition, a reduction of $n$ may increase barriers at the interface
\cite{Control_Schottky-barrier_height, Control_Schottky-barrier_height-1}; a usual Ni-silicide barrier ($\sim 0.7$~eV) may be completely restored at some domains or be still reduced ($\sim 0.5$~eV) at different places. Hole injection into the silicide layer from polysilicon grain boundaries may become more probable over reduced barriers to holes. 
This statement finds confirmation in the spectra plotted in Fig.\,\ref{fig:emf+IR} which have been obtained under irradiation of a diode by a wide-band IR radiation of a tungsten bulb filtered by a polished Si wafer [$h\nu < E_{\rm g}(300\,\mathrm{K}$)]. It is seen in the spectra that the higher the power density of the incident radiation  on the sample, the stronger the curves bow in the high-energy part of the graph and the less values of the photo-emf are detected.  It may be caused by injection of holes from potential wells at grain boundaries of poly-Si into the silicide film because of additional wide-band IR lighting of the sample resulting in charge reduction of both the silicide and polysilicon layers. 

Thus, a set of competing processes becomes possible at 80\,K. 
Non-uniformity of the spatial potential throughout the Ni-silicide/poly-Si interface may locally act in favour of one of these competing processes. As a consequence, impact of several barriers is observed in the photoresponse spectra in the order of magnitude of contribution of processes associated with them to the resultant photo-emf in different spectral ranges.
%!!!!!!!!!!!!!!!!!!!!!!!!!!

Investigating the temperature dependences of  the $I$--$V$ characteristics close and above the room temperature we have found the thermal sensitivity of the diodes  to be sufficiently high to consider them as a potential elements of uncooled bolometers. Panels (a) and (b) in Fig.\,\ref{fig:TSC-I} demonstrate temperature dependences  of the forward and reverse currents of the diodes  $I$ for fixed (and stabilized) voltages $U$. Temperature coefficient of the sensor current TCS\,$={\rm d}[\ln S(T)]/{\rm d}T$, where $S=I$, derived from the graphs presented in the panels (a) and (b) as a function of bias voltage (Fig.\,\ref{fig:TSC-I}\,c) vary from $-0.3$ to $-0.6$\,\%/{\textcelsius} for the forward bias and remains nearly constant around 2.5\,\%/{\textcelsius} for the reverse bias. Notice that at small values of the forward bias, TCS is positive but rapidly drops with the growth of the absolute bias   and equals 0 at  $U\approx-1$\,V. We think that the negative TCS may result from the metallic behavior of the poly-Si film as a function of temperature. Temperature dependences  of the  voltage drop across the diode $U$ for fixed (and stabilized) forward and reverse currents $I$ are shown in Fig.\,\ref{fig:TSC-U}\,a,\,b. 
The temperature coefficient of voltage TCS ($S=U$)  derived from the graphs depicted in Fig.\,\ref{fig:TSC-U}\,a,\,b (the curves in the panel (b) are linearized over an interval from 20 to 60\textcelsius)
vary from 0.3 to 0.6\,\%/{\textcelsius} for forward biasing and from $-3$ to $-2.4$\,\%/{\textcelsius} for reverse biasing (Fig.\,\ref{fig:TSC-U}\,c,\,d). 

As of now, we  foresee two ways of improvement of electrical properties of the structure. The first of them consists in modification of the Schottky barrier formation process proposed in Ref.\,\cite{Poly-Si_Schottky-diode} which enables production of poly-Si/Ni-polycide Schottky diodes with rectification ratios as high as $10^6$.  The other possibility is to replace poly-Si by $\alpha$-Si:H and to apply the metal induced crystallization to form diodes nearly as perfect as those produces on the basis of single-crystalline Si \cite{Ni-MIC-Si-review, Ni-MIC-Si-structural_properties, MIC-Si-p-n-junction, patent-Schottky-bolometer}. Each of these alternatives in principle could enable the development of high-performance monolithic  Schottky-diode  microbolometer IR FPAs.$^{\rm c}$

\section*{Conclusion}
In summary, nickel silicide Schottky diodes formed on polycrystalline Si$\langle$P$\rangle$ films are proposed as temperature sensors of monolithic uncooled microbolometer IR focal plane arrays. Structure and chemical composition of the Schottky diodes have been examined by TEM. The Ni silicide has been identified as multi-phase mixture composed by 20 to 40\% of Ni$_3$Si, 30 to 60\% of Ni$_2$Si and 10 to 30\% of NiSi with probable minor content of NiSi$_2$ at the silicide/poly-Si interface. $I$--$V$ characteristics of the diodes studied at different temperatures demonstrate the rectification ratios varying from $\sim$\,20 to  $\sim$\,100 when the temperature changes from 70 to 22{\textcelsius} and exceeding 1000 at 80\,K. A barrier of $\sim 0.95$~eV has been found to control the photovoltage spectra at room temperature. Three barriers with approximate heights from 1.08 to 1.14\,eV, from 0.66 to 0.78 and from 0.48 to 0.54 eV have been observed in photo-emf spectra at 80\,K and associated with the Ni-silicide/poly-Si interface. Absolute values of temperature coefficients of voltage and current have been found to vary from 0.3 to 0.6\,\%/{\textcelsius} for the forward biased structures and around 2.5\,\%/{\textcelsius} for the reverse biased ones.

%\cite{koon,oreg,khar,zvai,xjon,schn,pond,smith,marg,hunn,advi,koha,mouse}

%\section*{Section title}
%\subsection*{Sub-heading for section}
%\subsubsection*{Sub-sub heading for section}
%\subsubsection*{Sub-sub-sub heading for section}

\bigskip

%%%%%%%%%%%%%%%%%%%%%%%%%%%%%%%%

\section*{Abbreviations}
CMOS, complementary metal-oxide semiconductor; 
CZ, Czochralski or grown by the Czochralski method; 
emf, electromotive force;
FPA, focal plane array;
IR, infrared;
NETD, noise equivalent temperature difference;
TEM, transmission electron microscopy.

\section*{Competing interests}
The authors declare that they have no competing interests.

\section*{Author's contributions}
KVC participated in the design of the study, carried out the experiments, performed data analysis and participated in the discussions and interpretation of the results.
VAC participated in the design of the study, took part in the discussions and interpretation of the results; he also supervised the researches performed by young scientists and students.
VPK participated in the design of the study, took part in the discussions and interpretation of the results.
VYR performed the TEM studies and took part in the discussions and interpretation of the results.
MSS investigated the photo-emf spectra; he carried out the experiments, performed data analysis and took part in discussions and interpretation of the results.
VAY conceived of the study and designed it, performed data analysis and took part in the discussions and interpretation of the results; he also supervised the research project.
All authors read and approved the final manuscript.

\section*{Authors' information} 
KVC is a junior research fellow, VAC is a leading research fellow and MSS is a PhD student at Laboratory of Nanophotonics, Department of Applied Thermography, Prokhorov General Physics Institute, Russian Academy of Sciences.
VYR is a senior research fellow  and 
VPK is a head of Laboratory of Medium IR-range Crystalline Lasers at Department of Applied Thermography, Prokhorov General Physics Institute; VPK is also  a co-founder and a board member of  Technopark of GPI RAS and a co-founder and a partner in Thermographic Systems Ltd.
VAY is a  head of Department of Applied Thermography  and a head of Laboratory of Nanophotonics at Prokhorov General Physics Institute; he is also a co-founder and a board member of  Technopark of GPI RAS and a co-founder and a partner in Thermographic Systems Ltd.

%%%%%%%%%%%%%%%%%%%%%%%%%%%
\section*{Acknowledgements}
  \ifthenelse{\boolean{publ}}{\small}{}
  The equipment of the Center for Collective Use of Scientific Equipment of GPI RAS was used for this study. We acknowledge the technological support of our work.
We thank Ms.~N.~V.~Kiryanova for her valuable contribution to arrangement and management of this research. We express our appreciation to Mr.~V.~P.~Korol'kov and Mr.~G.~A.~Rudakov for performing the technological processes. We are grateful to Ms.~L.~A.~Krylova for carrying out chemical treatments of the experimental samples.

\section*{Endnotes}
%\footnote
$^{\rm a}${We cannot discriminate between $\delta$ and $\theta$ phases of Ni$_2$Si\cite{Silicides} and, following Ref.\,\cite{Murarka}, suppose that only the $\delta$ phase is present; the experimental value of its density, taken from Ref.\,\cite{Silicides}, makes 7.23~g/cm$^3$ whereas its X-ray density (7.405~g/cm$^3$) coincides in various sources \cite{Murarka, Silicides}.}%\pb

$^{\rm b}${A  barrier of this height is attributed   to the Ni/Si interface in Ref.\,\cite{Sze}; yet, we have not observed a direct contact of Ni to Si by TEM after the silicide film formation.}%\pb

$^{\rm c}$Notice also that there is an additional advantage of the considered structures with Schottky barriers: They may be applied both as temperature sensors of bolometers for detection in mid-IR or far-IR and as photonic sensors for detection in near-IR and visible spectral   ranges.
 
%%  Bmc_article.bst  will be used to                       %%
%%  create a .BBL file for submission, which includes      %%
%%  XML structured for BMC.                                %%
%%  After submission of the .TEX file,                     %%
%%  you will be prompted to submit your .BBL file.         %%

\newpage
{\ifthenelse{\boolean{publ}}{\footnotesize}{\small}
 \bibliographystyle{bmc_article}  % Style BST file
  \bibliography{NiSi_thermal_diodes} } % Bibliography file (usually '*.bib' ) 

%%%%%%%%%%%

\ifthenelse{\boolean{publ}}{\end{multicols}}{}

%%%%%%%%%%%%%%%%%%%%%%%%%%%%%%%%%%%
%%                               %%
%% Figures                       %%
%%                               %%
%% NB: this is for captions and  %%
%% Titles. All graphics must be  %%
%% submitted separately and NOT  %%
%% included in the Tex document  %%
%%                               %%
%%%%%%%%%%%%%%%%%%%%%%%%%%%%%%%%%%%

%%
%% Do not use \listoffigures as most will included as separate files

\section*{Figures}

  \subsection*{Figure 1 - TEM images demonstrate a Schottky-diode film  composed by three layers on Si$_3$N$_4$:}
 %     A short description of the figure content
  %    should go here.
(1) the Si$_3$N$_4$ substrate film;  
the diode film consists of 
(2) poly-Si , 
(3) nickel silicide  and 
(4) Al contact layers.

  \subsection*{Figure 2 - $I$--$V$ characteristics of the Ni-silicide/poly-Si structure and its rectification ratios at different temperatures: }
     % Figure legend text.
(a) forward  and 
(b) reverse biasing;
(c) rectification ratio vs the applied voltage.

 \subsection*{Figure 3 - $I$--$V$ characteristics of the Ni-silicide/poly-Si structure its rectification ratios at liquid nitrogen and room temperatures: }
     % Figure legend text.
(a) $I$--$V$ characteristics  and 
(b) rectification ratio as a function of the applied bias.

 \subsection*{Figure 4 - Photovoltage spectra obtained at room and liquid nitrogen temperatures:}
     % Figure legend text.
$\Phi_{\rm rt}$ is an estimate of the barrier height derived from the photo-emf spectral measurements at 300\,K with and without a Si filter;
$\Phi_0$, $\Phi_1$ and $\Phi_2$ are barrier heights estimated from the photo-emf spectral response at 80\,K (solid lines show line fits, dotted ones set 95\,\% confidence bands).

 \subsection*{Figure 5 - Photovoltage spectra obtained at 80\,K; the diode is irradiated by lite of a tungsten lamp through a Si filter:}
     % Figure legend text.
the power density of light with $h\nu < E_{\rm g}(300\,{\rm K})$ on the diode is shown in the legend in mW/cm$^2$; dashed lines are guides to the eye.

 \subsection*{Figure 6 - Temperature dependences of current for fixed voltages on a Ni-silicide/poly-Si Schottky  diode  and temperature coefficient of signal (current) for each branch of the {\it I--V} characteristics:}
% Figure legend text.
(a) forward  and 
(b) reverse  currents
 (the legend represents the applied bias in volts for each line); 
(c) temperature coefficient of current  vs fixed voltage on the structure;  negative and positive values of  $U$ in panel (c) correspond to forward  and  reverse biasing, respectively.

 \subsection*{Figure 7 - Temperature dependences of  voltage for fixed currents through a Ni-silicide/poly-Si Schottky  diode and temperature coefficient of signal (voltage) for each branch of {\it I--V} characteristics:}
% Figure legend text.
(a) forward and (b) reverse biases
(the legends represent the currents in $\mu$A for each line); 
(c, \,d) temperature coefficient of voltage for each branch of {\it I--V} characteristics  vs fixed current through the structure; 
to derive the graph (d), the curves in the panel (b) were linearized in the interval from 20 to 60{\textcelsius}; negative and positive values of $I$ in panels (c) and (d) correspond to forward  and  reverse biasing, respectively.

\newpage

\begin{figure}[h]
%Fig 1
\includegraphics[scale=0.45]{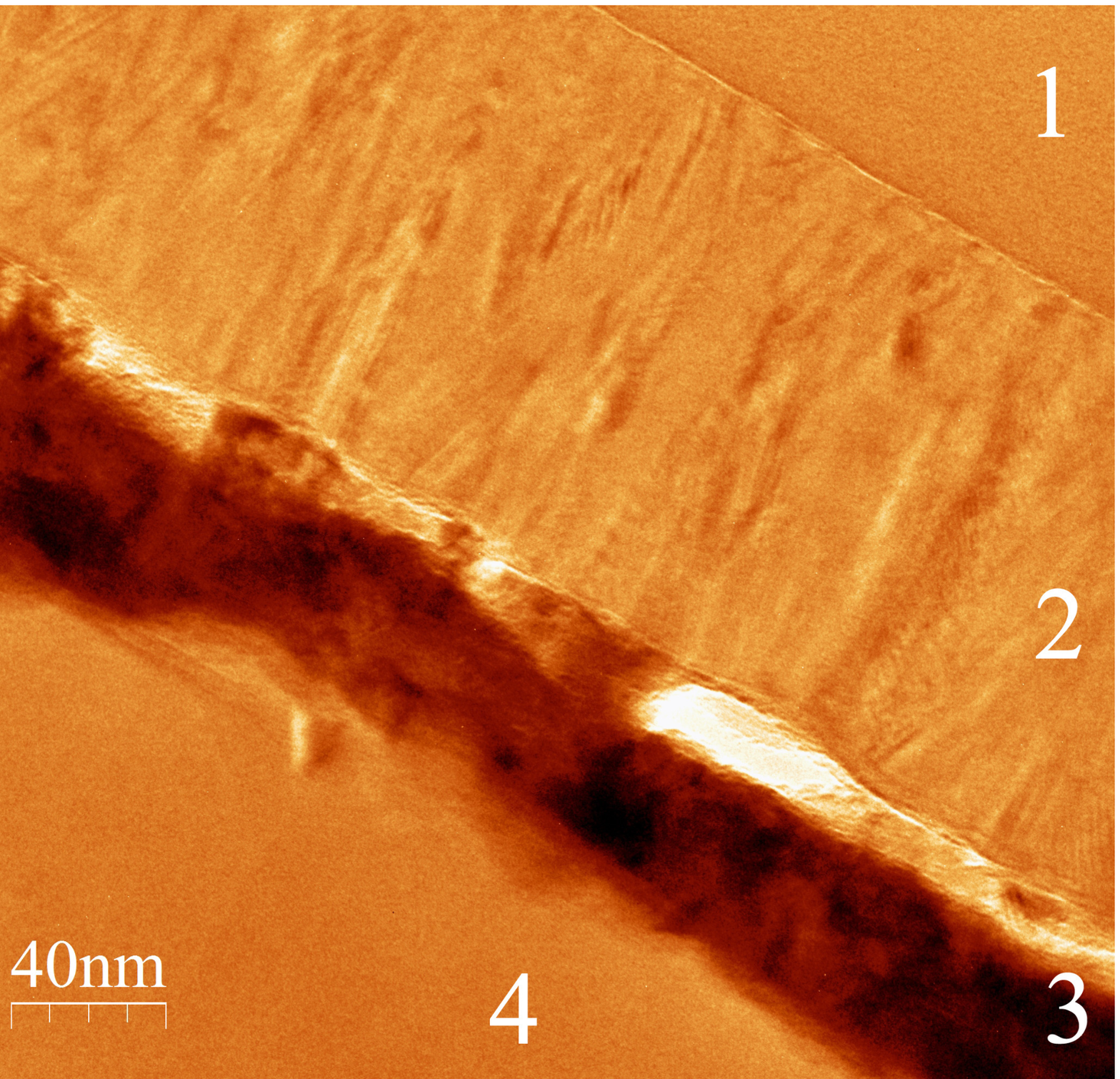}{\large(a)}
\includegraphics[scale=0.92]{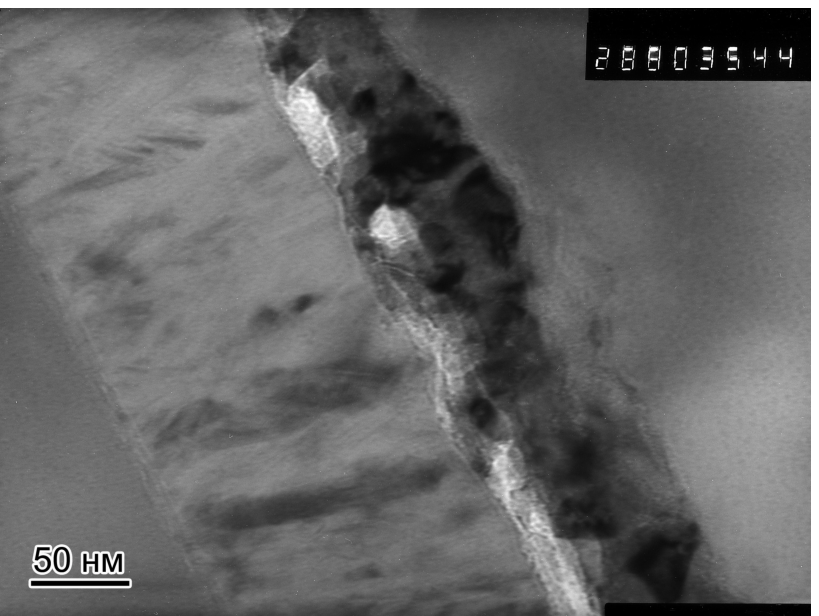}{\large(b)}
\caption{\label{fig:TEM}
TEM images demonstrate a Schottky-diode film  composed by three layers on Si$_3$N$_4$:
(1) the Si$_3$N$_4$ substrate film;  
the diode film consists of 
(2) poly-Si , 
(3) nickel silicide  and 
(4) Al contact layers.
}
\end{figure}

\begin{figure}[h]
%Fig 2
\includegraphics[scale=.4]{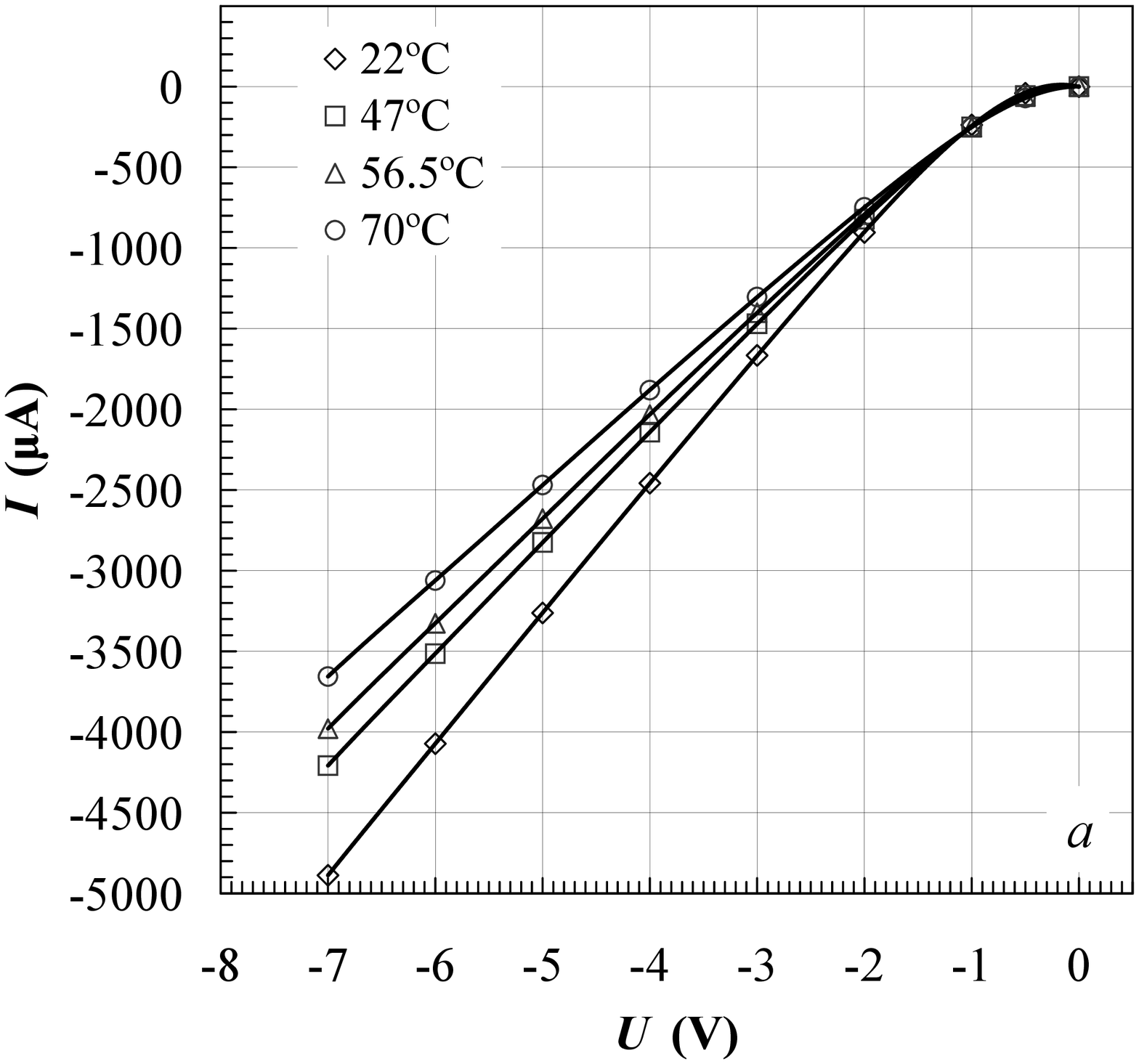}~~~%(a)
\includegraphics[scale=.4]{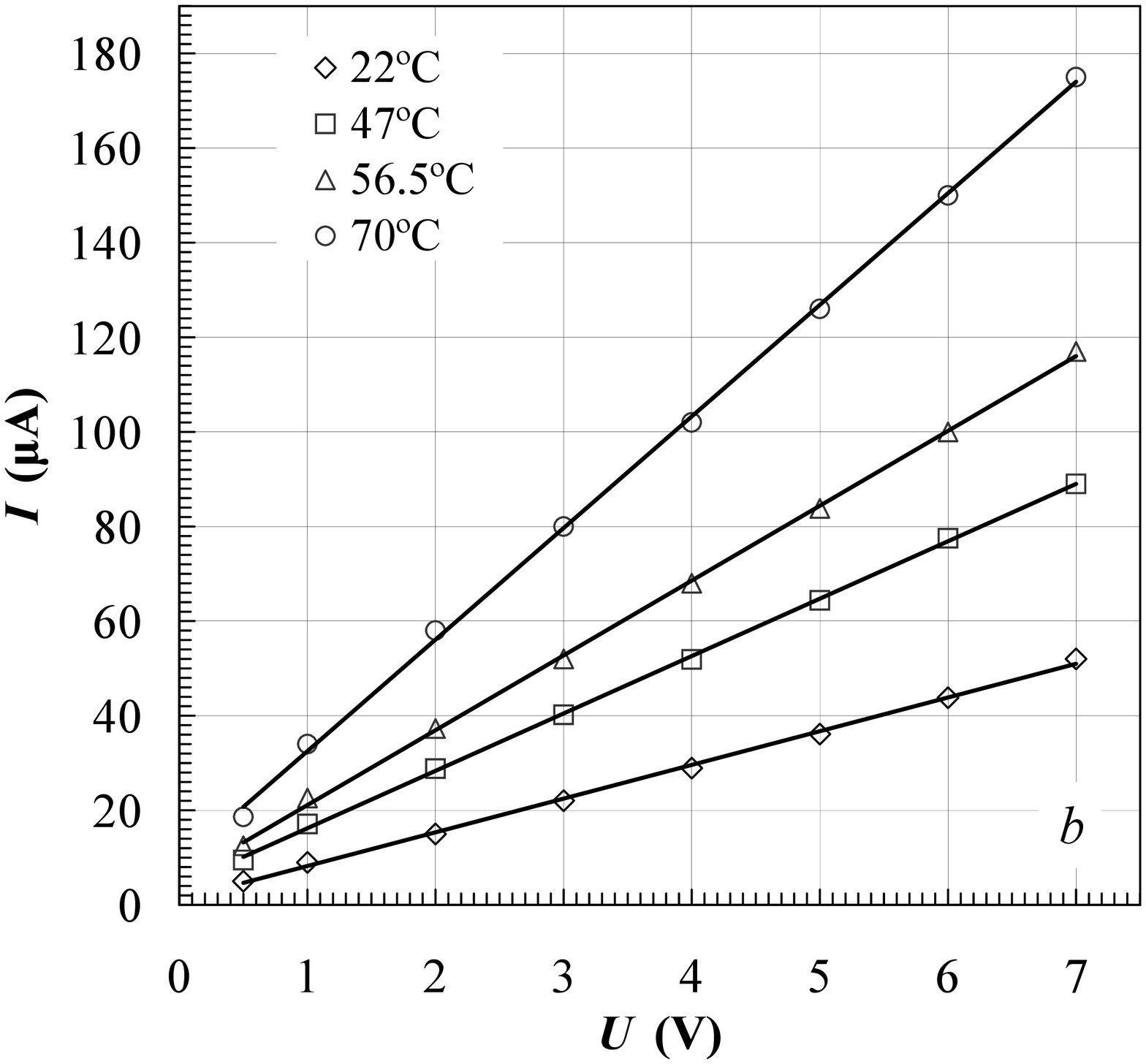}\\%(b)\\
\\
\includegraphics[scale=.4]{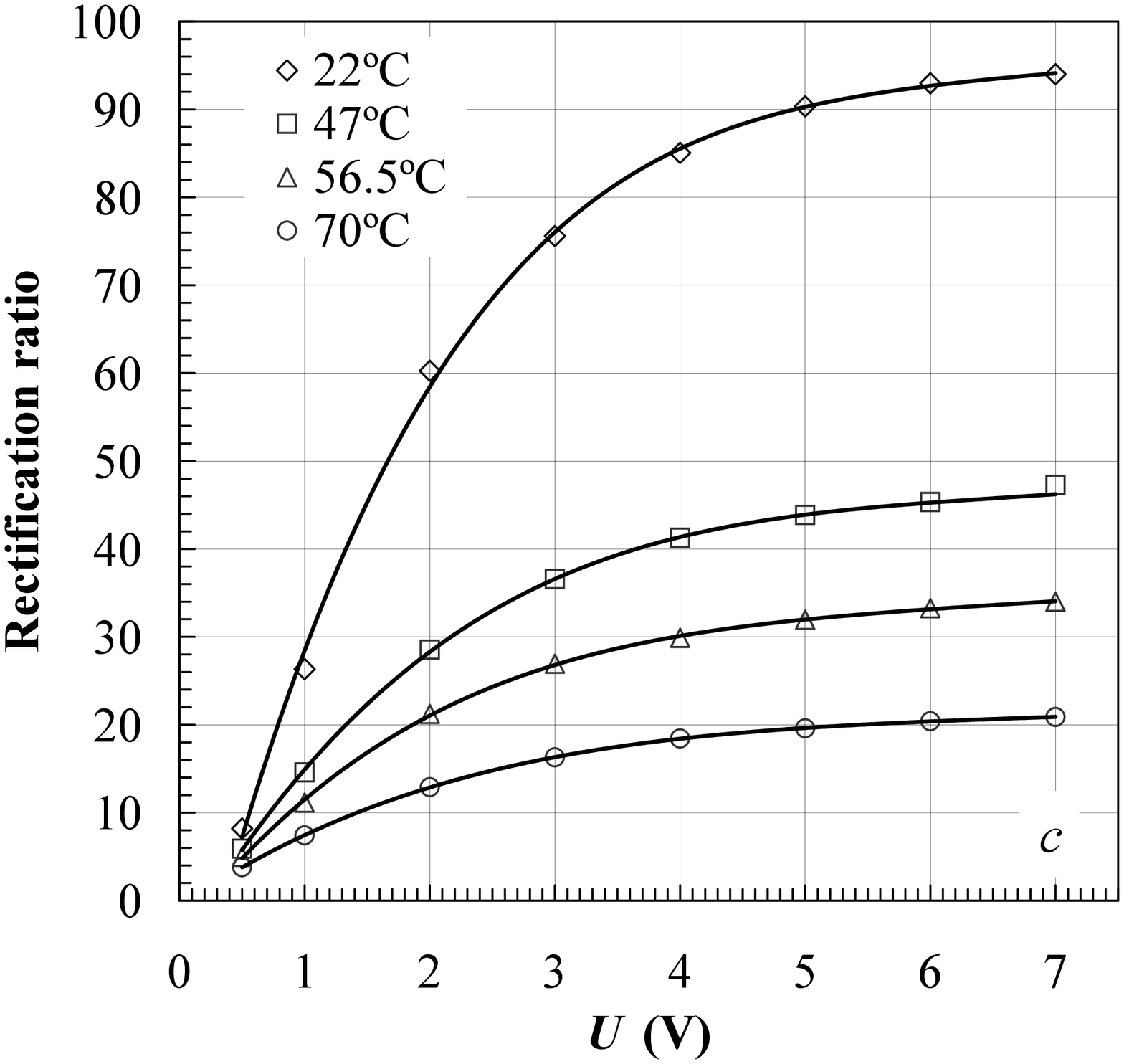}
\caption{\label{fig:I-V-300K}
$I$--$V$ characteristics of the Ni-silicide/poly-Si structure and its rectification ratios at different temperatures:  
(a) forward  and 
(b) reverse biasing;
(c) rectification ratio vs the applied voltage.
}
\end{figure}

\begin{figure}[h]
%Fig 3
\includegraphics[scale=.4]{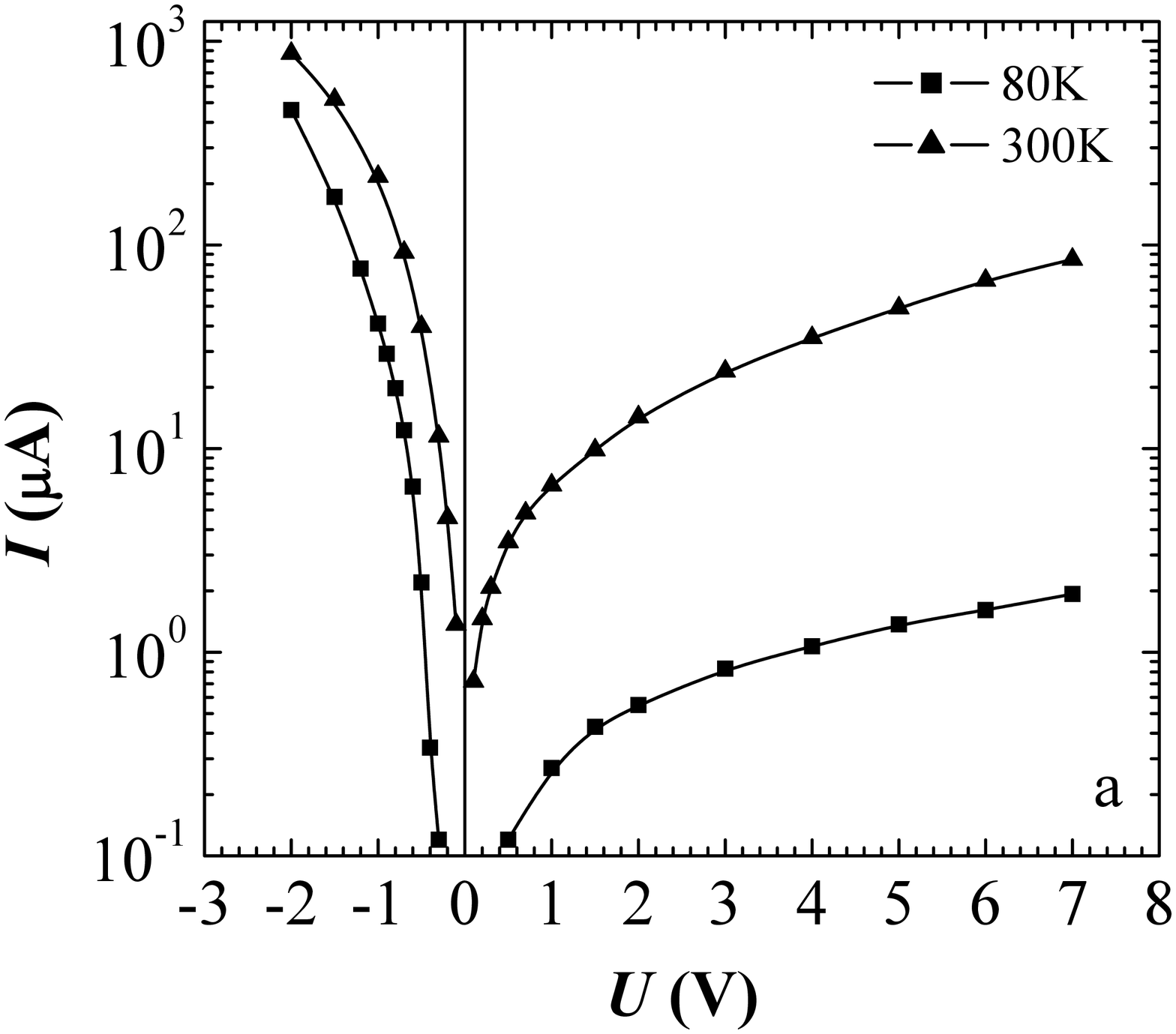}~~~%(a)
\includegraphics[scale=.415]{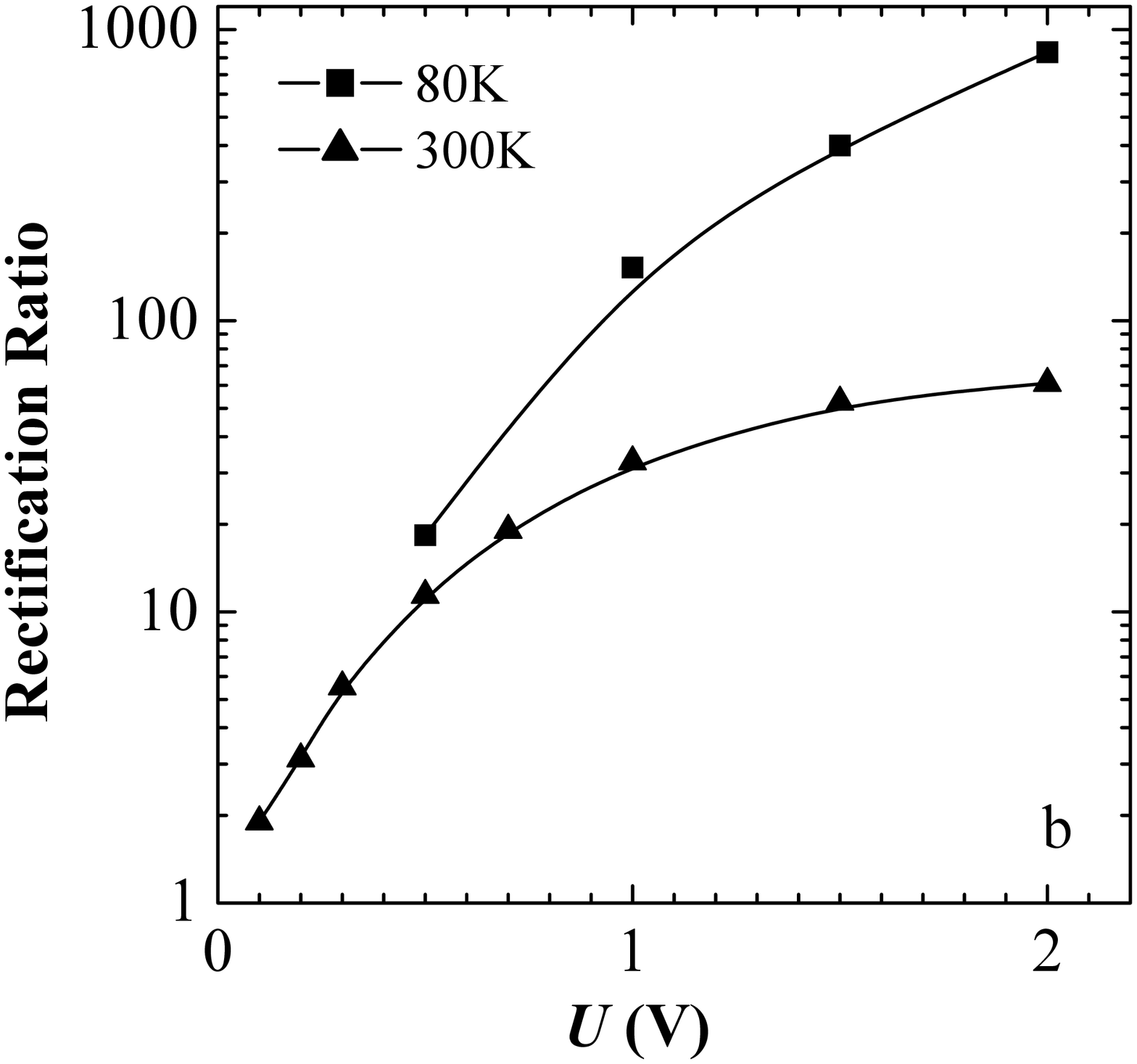}%(b)\\
\caption{\label{fig:I-V-80K}
$I$--$V$ characteristics of the Ni-silicide/poly-Si structure its rectification ratios at liquid nitrogen and room temperatures:  
(a) $I$--$V$ characteristics  and 
(b) rectification ratio as a function of the applied bias.
}
\end{figure}

\begin{figure}[h]
%Fig 4
\includegraphics[scale=.6]{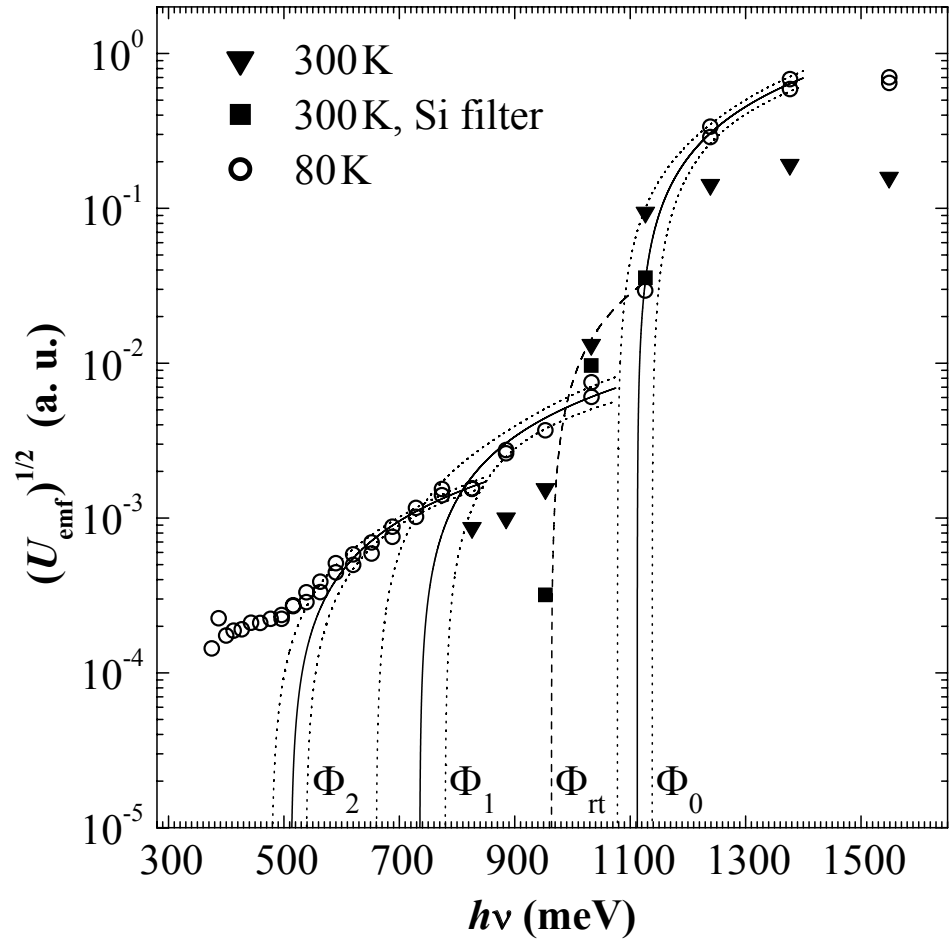}
\caption{\label{fig:emf}
Photovoltage spectra obtained at room and liquid nitrogen temperatures:
$\Phi_{\rm rt}$ is an estimate of the barrier height derived from the photo-emf spectral measurements at 300\,K with and without a Si filter;
$\Phi_0$, $\Phi_1$ and $\Phi_2$ are barrier heights estimated from the photo-emf spectral response at 80\,K (solid lines show line fits, dotted ones set 95\,\% confidence bands).
}
\end{figure}

\begin{figure}[h]
%Fig 5
\includegraphics[scale=.6]{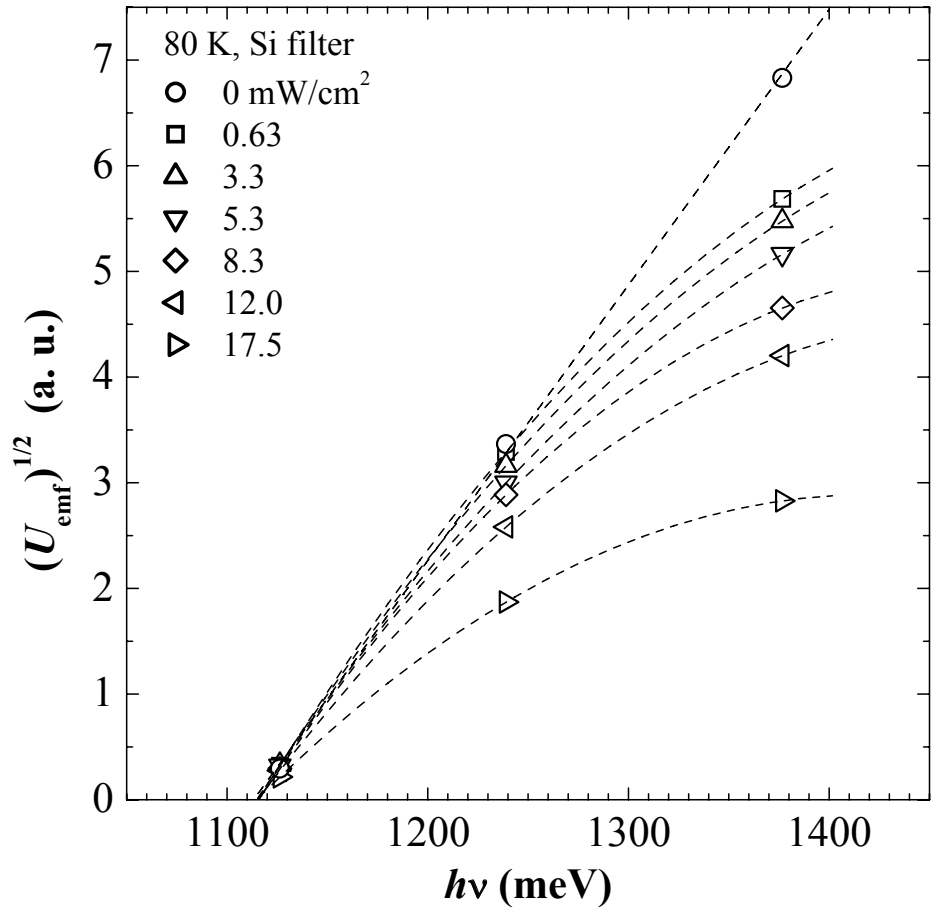}
\caption{\label{fig:emf+IR}
Photovoltage spectra obtained at 80\,K; the diode is irradiated by lite of a tungsten lamp through a Si filter:
the power density of light with $h\nu < E_{\rm g}(300\,{\rm K})$ on the diode is shown in the legend in mW/cm$^2$; dashed lines are guides to the eye.
}
\end{figure}

\begin{figure}[h]
%Fig 6
\includegraphics[scale=.4]{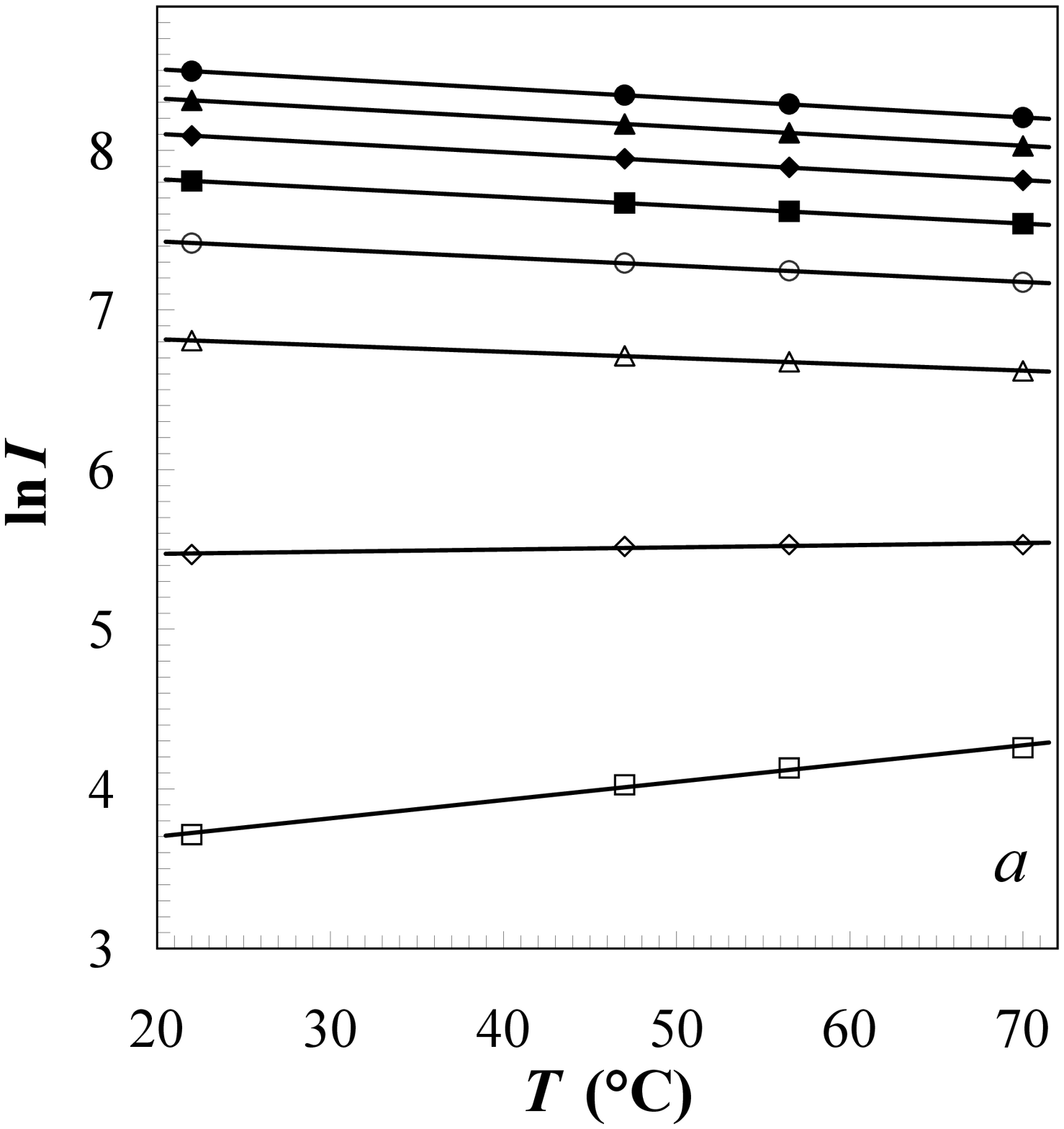}~~~%(a)
\includegraphics[scale=.4]{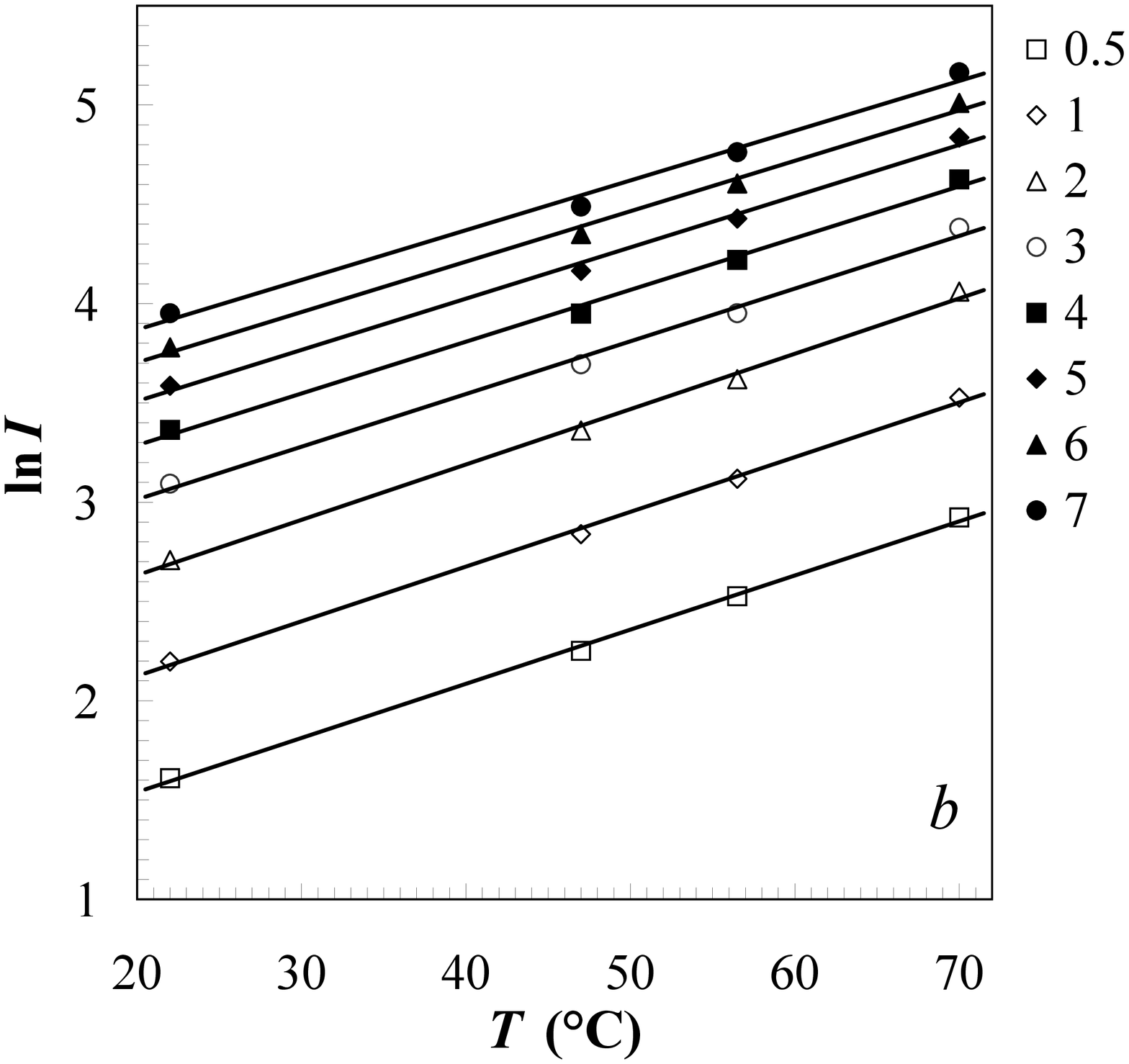}\\%(b)
\\
\includegraphics[scale=.4]{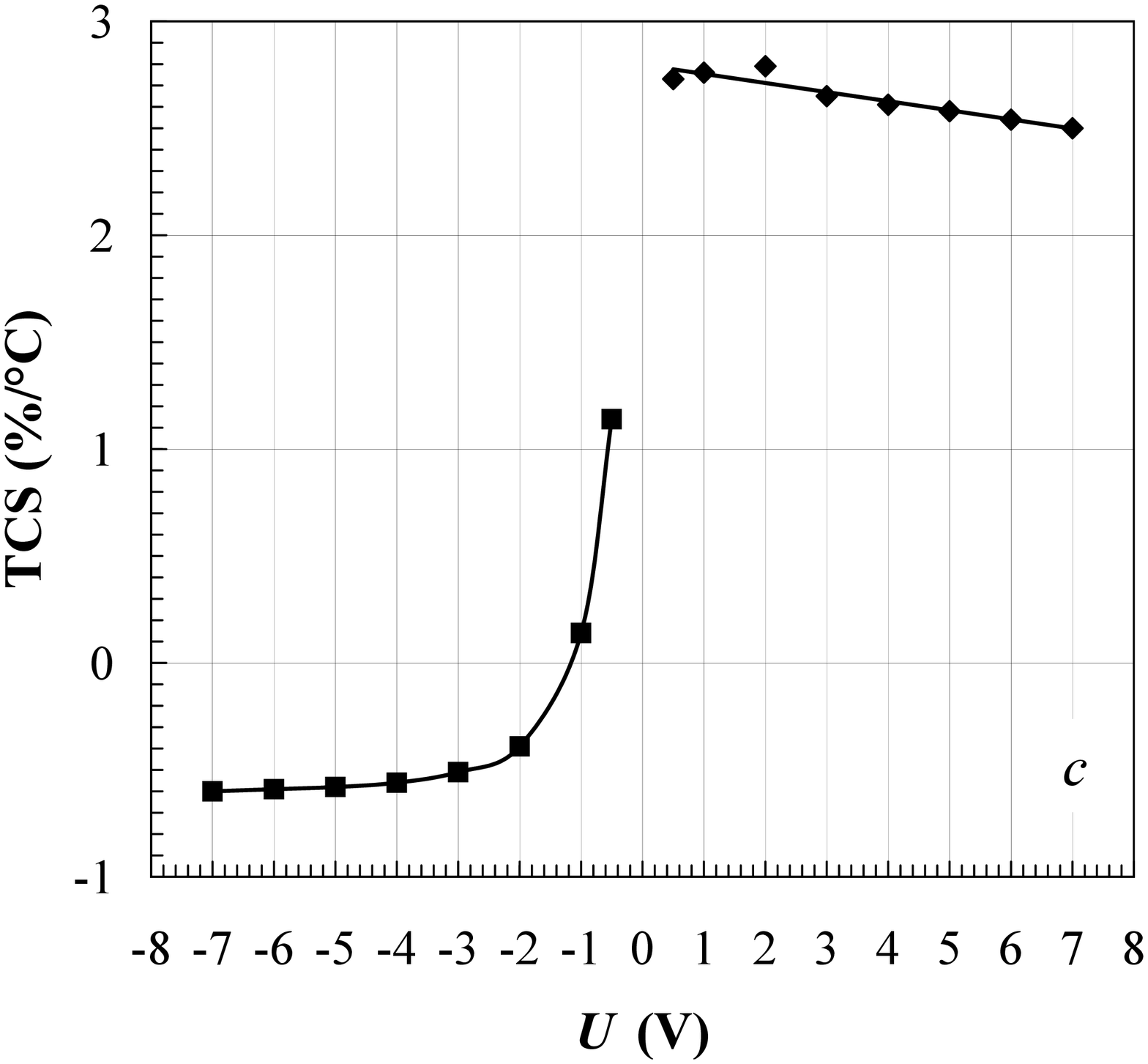}
\caption{\label{fig:TSC-I}
Temperature dependences of current for fixed voltages on a Ni-silicide/poly-Si Schottky  diode  and temperature coefficient of signal (current) for each branch of the {\it I--V} characteristics:
(a) forward  and 
(b) reverse  currents
 (the legend represents the applied bias in volts for each line); 
(c) temperature coefficient of current  vs fixed voltage on the structure;  negative and positive values of $U$ in panel (c) correspond to forward  and  reverse biasing, respectively.
}
\end{figure}

\begin{figure}[h]
%Fig 7
\includegraphics[scale=.42]{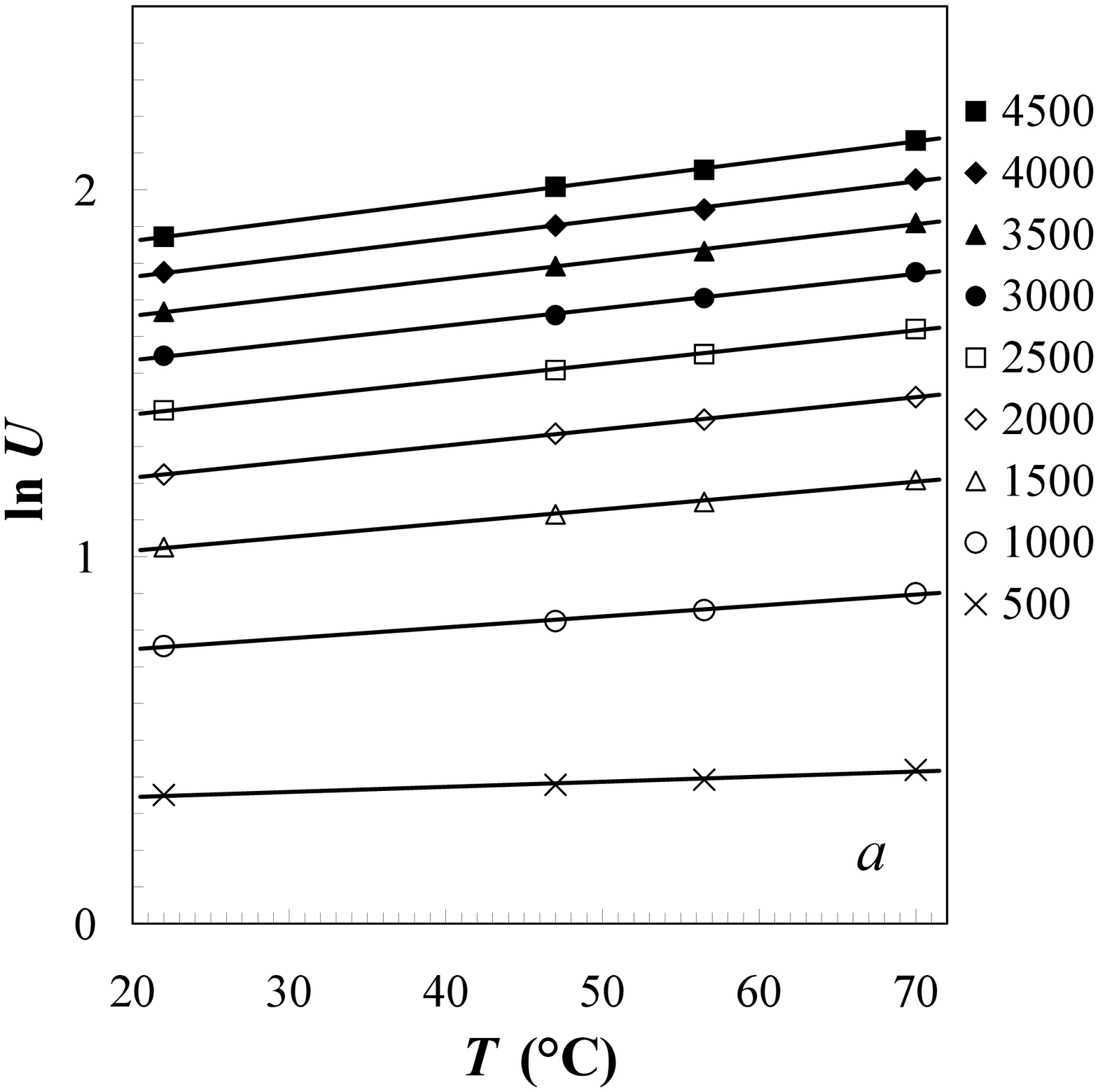}~~~%(a)
\includegraphics[scale=.42]{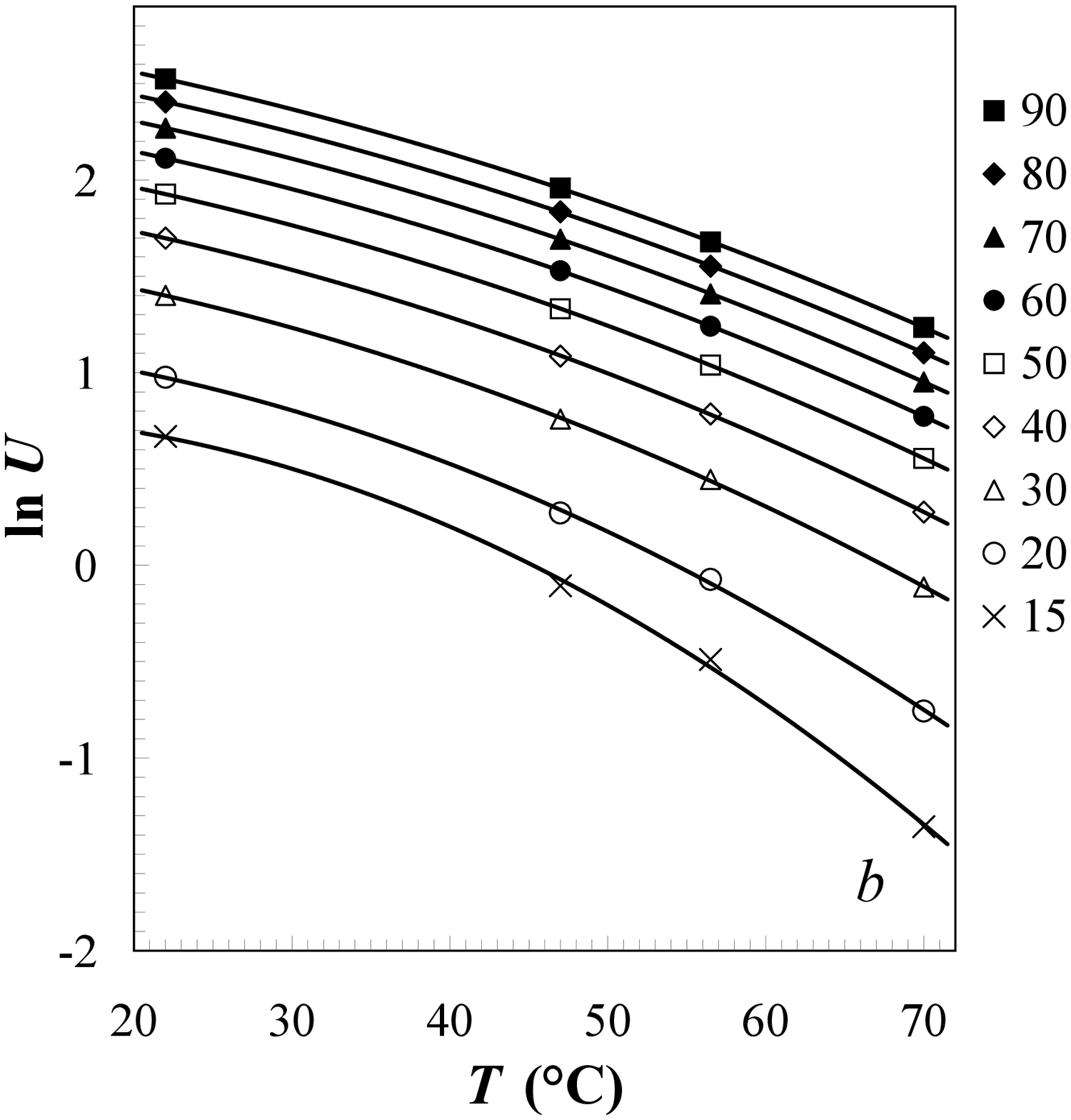}\\%(b)
\\
\includegraphics[scale=.4]{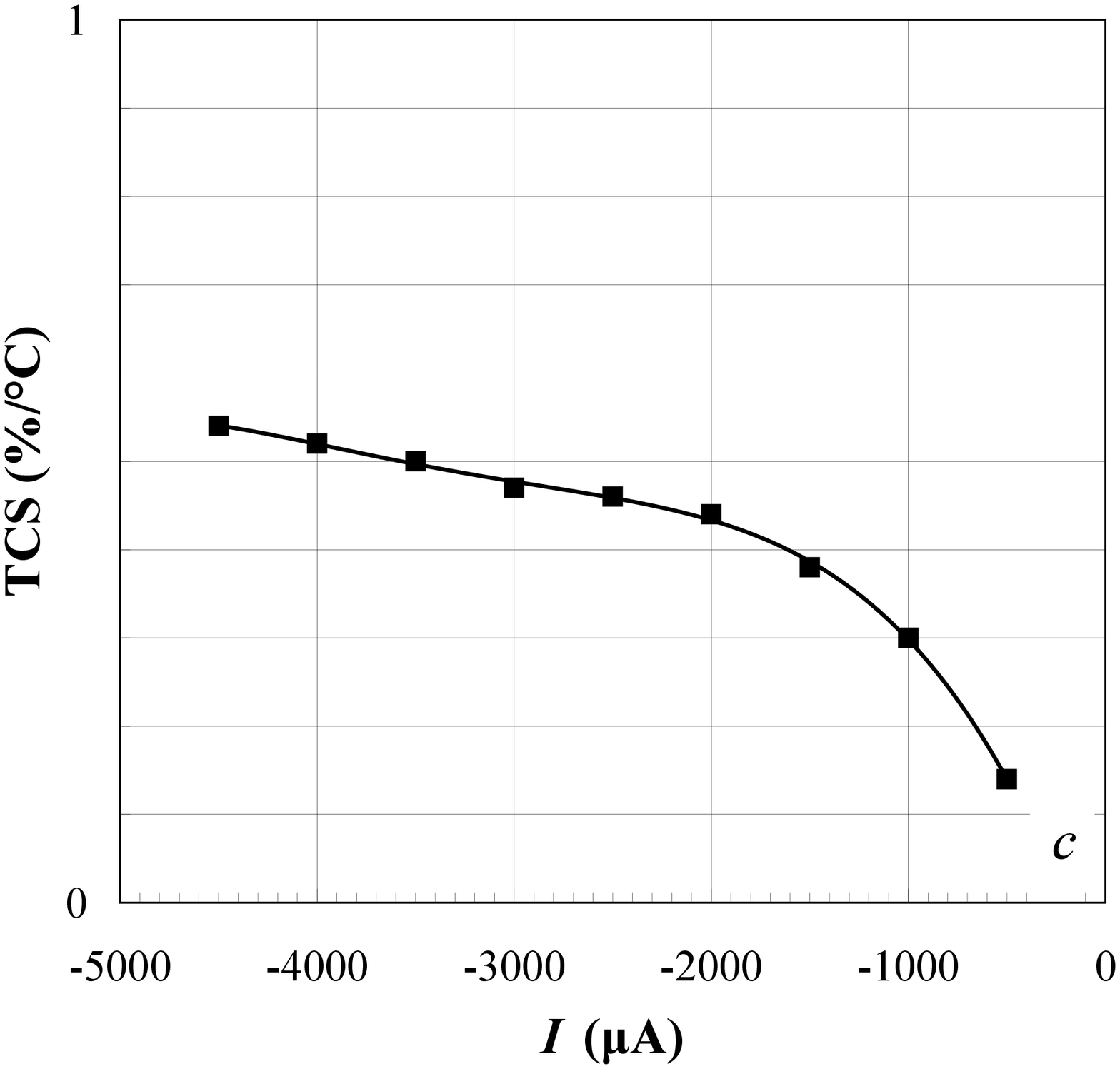}~~%(c)
\includegraphics[scale=.4]{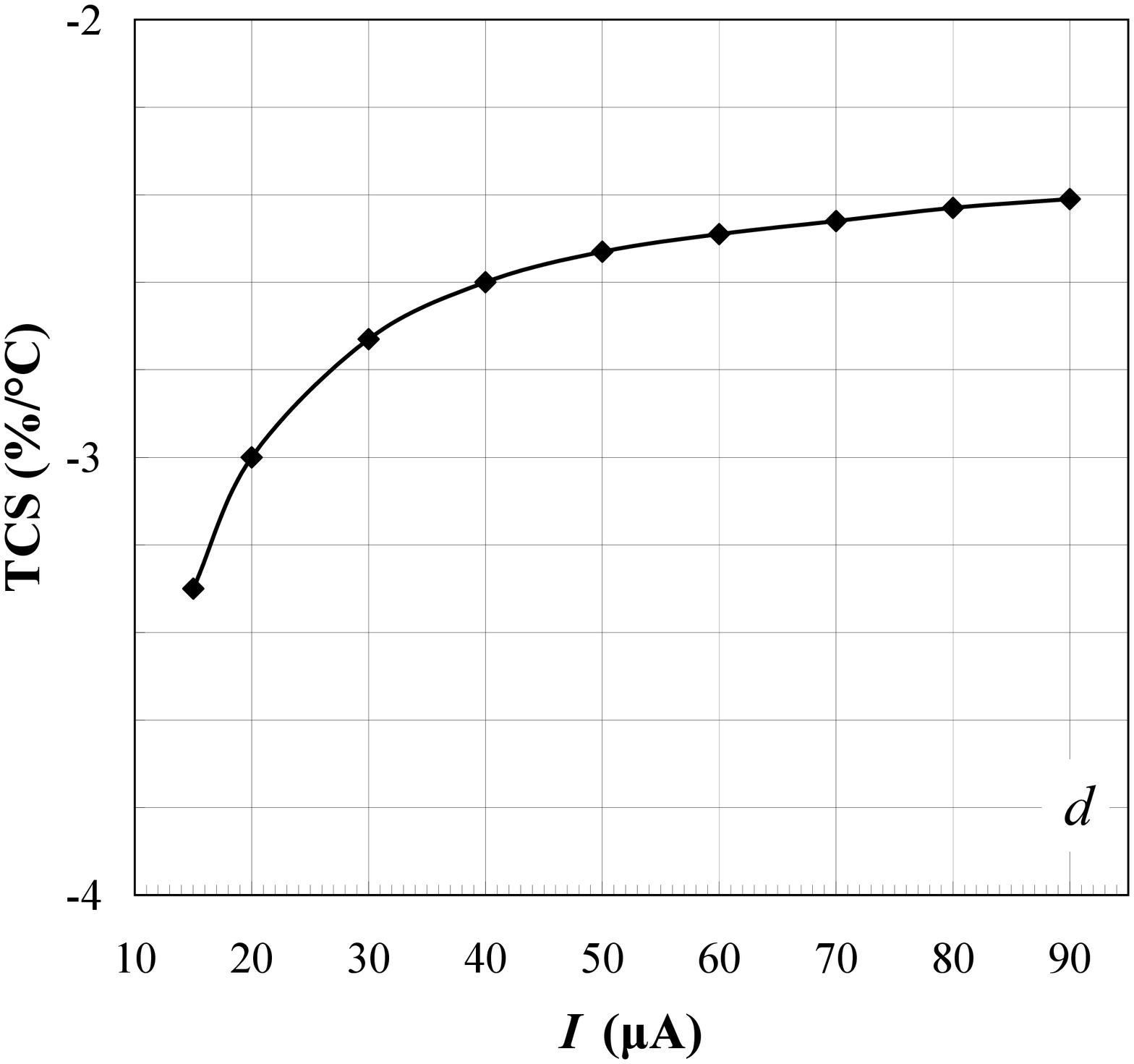}%(d)
\caption{\label{fig:TSC-U}
Temperature dependences of  voltage for fixed currents through a Ni-silicide/poly-Si Schottky  diode and temperature coefficient of signal (voltage) for each branch of {\it I--V} characteristics:
(a) forward and (b) reverse biases
(the legends represent the currents in $\mu$A for each line); 
(c, \,d) temperature coefficient of voltage for each branch of {\it I--V} characteristics  vs fixed current through the structure; 
to derive the graph (d), the curves in the panel (b) were linearized in the interval from 20 to 60{\textcelsius}; negative and positive values of $I$ in panels (c) and (d) correspond to forward  and  reverse biasing, respectively.
}
\end{figure}

%%%%%%%%%%%%%%%%%%%%%%%%%%%%%%%%%%%
%%                               %%
%% Tables                        %%
%%                               %%
%%%%%%%%%%%%%%%%%%%%%%%%%%%%%%%%%%%

%% Use of \listoftables is discouraged.
%%
%\section*{Tables}
 % \subsection*{Table 1 - Sample table title}
%    Here is an example of a \emph{small} table in \LaTeX\ using  
%    \verb|\tabular{...}|. This is where the description of the table 
 %   should go. \par \mbox{}
 %   \par
 %   \mbox{
 %     \begin{tabular}{|c|c|c|}
 %       \hline \multicolumn{3}{|c|}{My Table}\\ \hline
 %       A1 & B2  & C3 \\ \hline
 %       A2 & ... & .. \\ \hline
 %       A3 & ..  & .  \\ \hline
 %     \end{tabular}
%      }

%%%%%%%%%%%%%%%%%%%%%%%%%%%%%%%%%%%
%%                               %%
%% Additional Files              %%
%%                               %%
%%%%%%%%%%%%%%%%%%%%%%%%%%%%%%%%%%%

%\section*{Additional Files}
  %\subsection*{Additional file 1 --- Sample additional file title}
  %  Additional file descriptions text (including details of how to
 %   view the file, if it is in a non-standard format or the file extension).  This might
 %   refer to a multi-page table or a figure.

%  \subsection*{Additional file 2 --- Sample additional file title}
 %   Additional file descriptions text.

\end{bmcformat}
\end{document}